\documentclass[preprintnumbers,amsmath,amssymb,twocolumn,nopacs,prl]{revtex4}
\usepackage{graphicx}% Include figure files
\usepackage{dcolumn}% Align table columns on decimal point
\usepackage{bm}% bold math
\usepackage{color}
\usepackage{mathtools}

\usepackage{hyperref}

\graphicspath{{Figures/}}

\begin{document}

\title{Room temperature skyrmions at zero field in exchange-biased ultrathin films}
\author{K. Gaurav Rana$^{1}$\footnote{These authors contributed equally to this work}, A.~Finco$^{2}$\footnotemark[\value{footnote}], F. Fabre$^{2}$, S.~Chouaieb$^{2}$, A. Haykal$^{2}$, L.~D.~Buda-Prejbeanu$^{1}$, O.~Fruchart$^{1}$, S. Le Denmat$^{3}$, P.~David$^{3}$, M.~Belmeguenai$^{4}$, T. Denneulin$^{5}$, R. E. Dunin-Borkowski$^{5}$, G. Gaudin$^{1}$, V.~Jacques$^{2}$ and O. Boulle$^{1}$}
\affiliation{$^{1}$Univ. Grenoble Alpes, CNRS, CEA, Grenoble INP, Spintec, Grenoble, France}
\affiliation{$^{2}$Laboratoire Charles Coulomb, Universit\'e de Montpellier and CNRS, 34095 Montpellier, France}
\affiliation{$^{3}$Univ. Grenoble  Alpes, CNRS,  Institut  N\'eel,  38000  Grenoble,  France}
\affiliation{$^{4}$LSPM, CNRS-Universit\'e Paris 13, Sorbonne Paris Cit\'e, 93430 Villetaneuse, France}
\affiliation{$^{5}$Ernst Ruska-Centre for Microscopy and Spectroscopy with Electrons and Peter Gr$\ddot{u}$nberg Institute, Forschungszentrum J$\ddot{u}$lich, 52425 J$\ddot{u}$lich, Germany }

\begin{abstract}
We demonstrate that magnetic skyrmions with a mean diameter around $60$~nm can be stabilized at room temperature and zero external magnetic field in an exchange-biased Pt/Co/NiFe/IrMn multilayer stack. This is achieved through an advanced optimization of the multilayer stack composition in order to balance the different magnetic energies controlling the skyrmion size and stability. Magnetic imaging is performed both with magnetic force microscopy and scanning Nitrogen-Vacancy magnetometry, the latter providing unambiguous measurements at zero external  magnetic field. In such samples, we show that exchange bias provides an immunity of the skyrmion spin texture to moderate external magnetic field, in the tens of mT range, which is an important feature for applications as memory devices. These results establish exchange-biased multilayer stacks as a promising platform towards the effective realization of memory and logic devices based on magnetic skyrmions.
\end{abstract}
\date{\today}

\maketitle

%%%%%%%%%%%%%%%%%%%%%%%%%%%INTRO%%%%%%%%%%%%%%%%%%%%%%%%%%%%%%%%%%%%%%%%%%%%%%%%%%%%%%%%%%%

Magnetic skyrmions are whirling spin textures, which hold great promise to store and process the information at the nanoscale in future memory and logic devices. These topological spin textures were first observed at low temperature and in the presence of large magnetic fields both in bulk materials exhibiting broken inversion symmetry~\cite{Muhlbauer915,yu2010} and in epitaxial ultrathin films with interfacial Dzyaloshinskii-Moriya interaction~\cite{heinze2011,romming2013}. After years of active research, magnetic skyrmions can nowadays be stabilized at room temperature in technologically-relevant materials based on sputtered magnetic multilayer systems lacking inversion symmetry~\cite{Jiang2015,woo2016,boulle2016,moreau-luchaire2016}. Deterministic skyrmion nucleation and fast current-induced motion ($>100$~m/s) were also recently demonstrated, owing to efficient spin orbit torques in such magnetic multilayer stacks~\cite{Fert2017,RevueSk}. These results have opened a path for memory and logic devices where skyrmions in tracks are used as information carriers. Their particle-like, topologically stable spin texture, small size and efficient current-induced manipulation could lead to a unique combination of high density, fast and high data flow capabilities~\cite{Fert2017,RevueSk}.

 To realize such devices, small-sized skyrmions must be stabilized at ambient conditions without the need of an external magnetic field. This challenging task can be pursued by using either confined  geometries~\cite{boulle2016,PhysRevApplied.11.024064}, metastable skyrmion lattice textures~\cite{doi:10.1063/1.5080713}, frustration of exchange interaction in ultrathin ferromagnets~\cite{Heinze_NatCom2019} or ferrimagnetic materials close to the magnetization compensation~\cite{Caretta2018}. Another promising strategy consists in designing magnetic heterostructures in which interlayer exchange coupling acts as an effective {\it internal} magnetic field $B_{\rm int}$~\cite{APL2015,NanoLett2018}. Here we demonstrate that magnetic skyrmions with a mean diameter around $60$~nm can be stabilized at room temperature and zero external magnetic field in an ultrathin ferromagnetic layer exchange-biased by an antiferromagnetic film.  Compared to previous studies using exchange-biased multilayer stacks~\cite{NanoLett2018}, this result corresponds to a reduction of skyrmion diameter by one-order of magnitude. This is achieved through an advanced optimization of the multilayer stack composition in order to balance the different magnetic energies controlling the skyrmion size and stability~\cite{buttner2018,Mantel2018}. 
 
 \begin{figure}[h!]
\includegraphics[width = 8.6 cm]{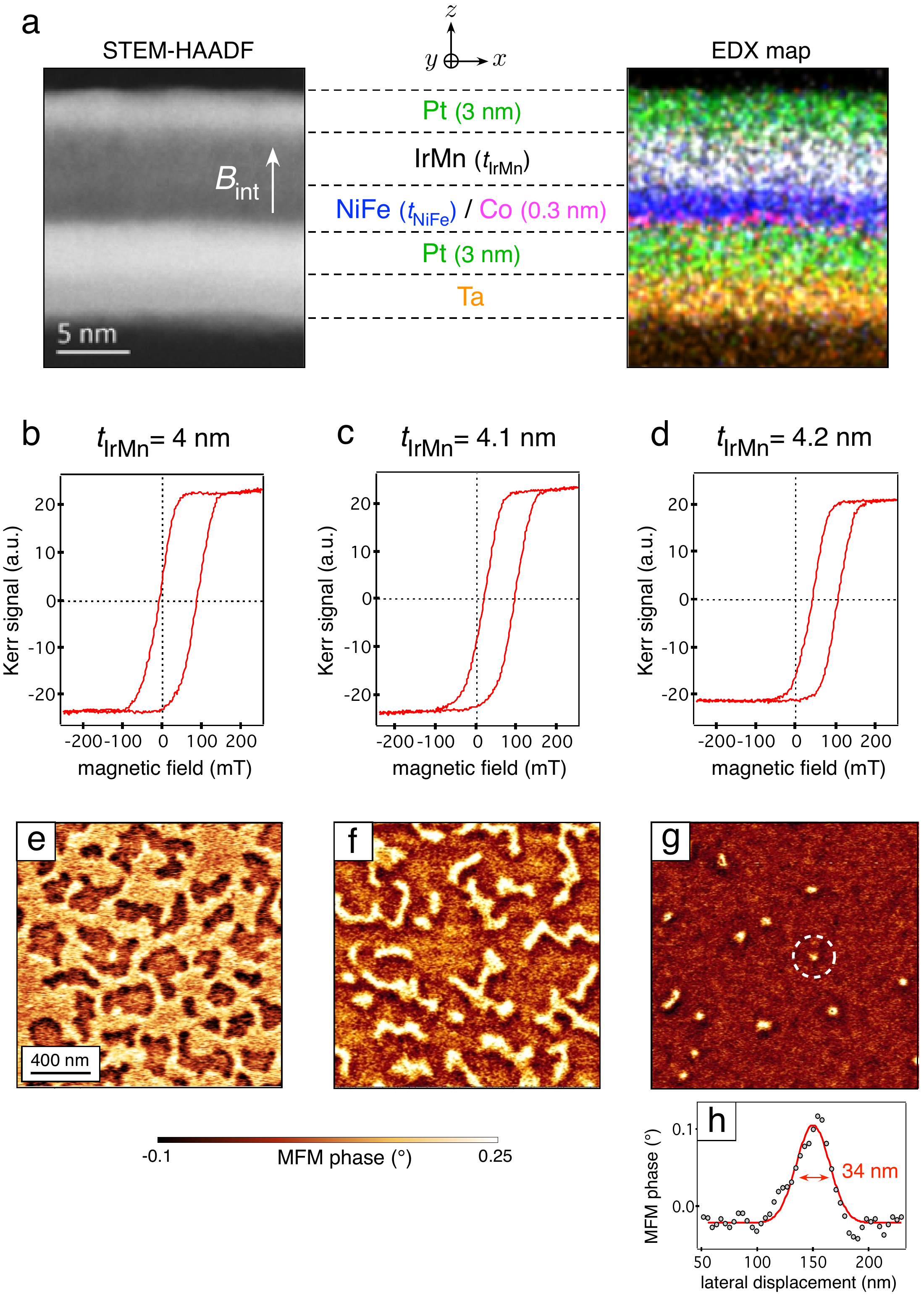}
\caption{{\bf (a)} Central panel: Schematic structure of the exchange-biased multilayer stack. Left panel: Scanning transmission electron microscopy image of a cross-section of the sample obtained in a high-angle annular dark-field imaging (STEM-HAADF) mode. Right panel: energy-dispersive X-ray (EDX) elemental map. The white arrow indicates the direction of the internal exchange bias field $B_{\rm int}$ across the NiFe/IrMn interface. {\bf (b-d)} Magneto-optical Kerr rotation recorded as a function of a magnetic field applied perpendicularly to the film plane for various IrMn thicknesses, (b) $t_{\rm IrMn}=4$~nm, (c) $t_{\rm IrMn}=4.1$~nm and (d) $t_{\rm IrMn}=4.2$~nm. For these experiments, the NiFe thickness is $t_{\rm NiFe}=2.1$~nm. {\bf (e-g)} Corresponding magnetic force microscopy (MFM) images recorded at zero external magnetic field. {\bf (h)} Line profile across the skyrmion marked with the white dashed circle in (g). The red solid line is a Gaussian fit, leading to a full width at half maximum $w_{\rm MFM}=34\pm3$~nm.}
\label{Fig1}
\end{figure}

The studied exchange-biased sample consists of a Ta/Pt(3)/Co(0.3)/Ni$_{80}$Fe$_{20}$($t_{\rm NiFe}$)/Ir$_{20}$Mn$_{80}$($t_{\rm IrMn}$)/Pt(3) multilayer stack (unit in nm) deposited by magnetron sputtering on a 100 mm diameter Si substrate [Fig.~1a].  The Pt/Co interface yields perpendicular magnetic anisotropy combined with a sizable interfacial Dzyaloshinskii-Moriya interaction (DMI)~\cite{KaiPRL2015,Belmeguenai2015}, which is a key ingredient for stabilizing skyrmions. Furthermore, this interface provides large spin orbit torques for efficient current induced skyrmion motion~\cite{PhysRevApplied.12.044007}. The use of a NiFe thin film offers (i) a large exchange bias at the interface with the IrMn layer and (ii) a lower spontaneous magnetization compared to cobalt, so that small-sized skyrmions are expected owing to a reduced dipolar energy~\cite{buttner2018,Mantel2018}. To induce exchange bias, the samples were annealed at 200$^{\circ}$C for 2 minutes under a large out-of-plane magnetic field $B_z=570$~mT. Element-resolved scanning transmission electron microscopy images of a cross-section of the sample reveal sharp interfaces with negligible intermixing [Fig.~1a]. 

An important parameter controlling the skyrmion size and stability is the NiFe film thickness, $t_{\rm NiFe}$, since the main skyrmion energy terms, namely the magnetic anisotropy, DMI and dipolar fields, are strongly dependent on the thickness of the ferromagnetic layer~\cite{boulle2016}. Similarly, the exchange bias and magnetic anisotropy are highly sensitive to the IrMn film thickness $t_{\rm IrMn}$~\cite{NanoLett2018}. In order to find the optimal sample composition stabilizing zero-field skyrmions, gradients of NiFe and IrMn thicknesses were deposited along the $x$ and $y$ axis of the 100-mm wafer, respectively, using off-axis deposition. Systematic hysteresis loop mapping was then carried out using magneto-optical Kerr effect supplemented by magnetic force microscopy (MFM). 

Optimal sample composition stabilizing zero-field skyrmions was achieved for a NiFe thickness around $t_{\rm NiFe}=2.1$~nm, close to the out-of-plane to in-plane transition of the magnetization, and an IrMn thickness around $t_{\rm IrMn}=4.2$~nm. Typical magnetic hysteresis loops recorded for different thicknesses of the IrMn layer are shown in Figure 1(b-d). The observation of bended hysteresis loops suggests the existence of multidomain states. In addition, these measurements indicate that the effective exchange bias field $B_{\rm int}$, characterized by the overall shift of the hysteresis loop, increases with the IrMn film thickness, while the coercivity decreases. MFM images recorded at zero external magnetic field for different IrMn thicknesses, {\it i.e.} for different exchange bias fields $B_{\rm int}$, are shown in Figure~1(e-g). For $t_{\rm IrMn}=4$~nm, the small exchange bias field combined with a large coercivity of the ferromagnetic layer leads to a multidomain state, with a nearly equal density of up- and down-magnetized domains [Fig.~1e]. As the IrMn thickness is slightly increased to $t_{\rm IrMn}=4.1$~nm, down-magnetized domains start to shrink and few skyrmions can be observed [Fig.~1f]. Further increase in IrMn thickness ($t_{\rm IrMn}=4.2$~nm) leads to the stabilization of well-isolated skyrmions in the ferromagnetic layer [Fig.~1g]. The smaller skyrmions exhibit a full width at half maximum down to $\sim35$~nm in the MFM image [Fig.~1h].   

In the following, we focus on the optimized Pt/Co(0.3)/NiFe(2.1)/IrMn(4.2) multilayer stack. Its effective saturation magnetization was measured by SQUID magnetometry, leading to $M_s=720$~kA/m, while vibrating sample magnetometry was used to infer the perpendicular magnetic anisotropy $K=350$~kJ/m$^3$. The strength of interfacial DMI was determined by monitoring the non-reciprocal propagation of spin waves with Brillouin light scattering (BLS) spectroscopy~\cite{KaiPRL2015,Belmeguenai2015,Nembach2015}, yielding $D=-0.4$~mJ/m$^2$. The negative value of the $D$ parameter promotes the stabilization of N\'eel skyrmions with a left-handed chirality.

A systematic study of the skyrmion size was realized through magnetic imaging with a scanning Nitrogen-Vacancy (NV) magnetometer operated in photoluminescence (PL) quenching mode at room temperature and zero external magnetic field~\cite{Rondin2014}. Compared to MFM, the main advantage of this technique is here the absence of magnetic perturbations on the studied sample, thus providing unambiguous measurements at zero magnetic field. We employ a single NV defect located at the apex of a nanopillar in a commercial diamond scanning-probe unit~\cite{Maletinsky2012,Appel} (Qnami, Quantilever MX), which is integrated into an atomic force microscope and scanned in close proximity to the magnetic sample. At each point of the scan, a confocal optical microscope is used to record the PL signal of the NV defect. Stray magnetic fields produced locally by spin textures in ferromagnets lead to an overall reduction of the PL signal~\cite{Tetienne2012,gross2018,PhysRevApplied.11.034066}. Such a magnetic-field-induced PL quenching is here exploited to image magnetic skyrmions with a spatial resolution fixed by the distance between the NV sensor and the ferromagnetic layer, $z_{\rm NV}\approx 65$~nm, as measured through an independent calibration procedure~\cite{Hingant2015}.

\begin{figure}[t]
\includegraphics[width = 8.7cm]{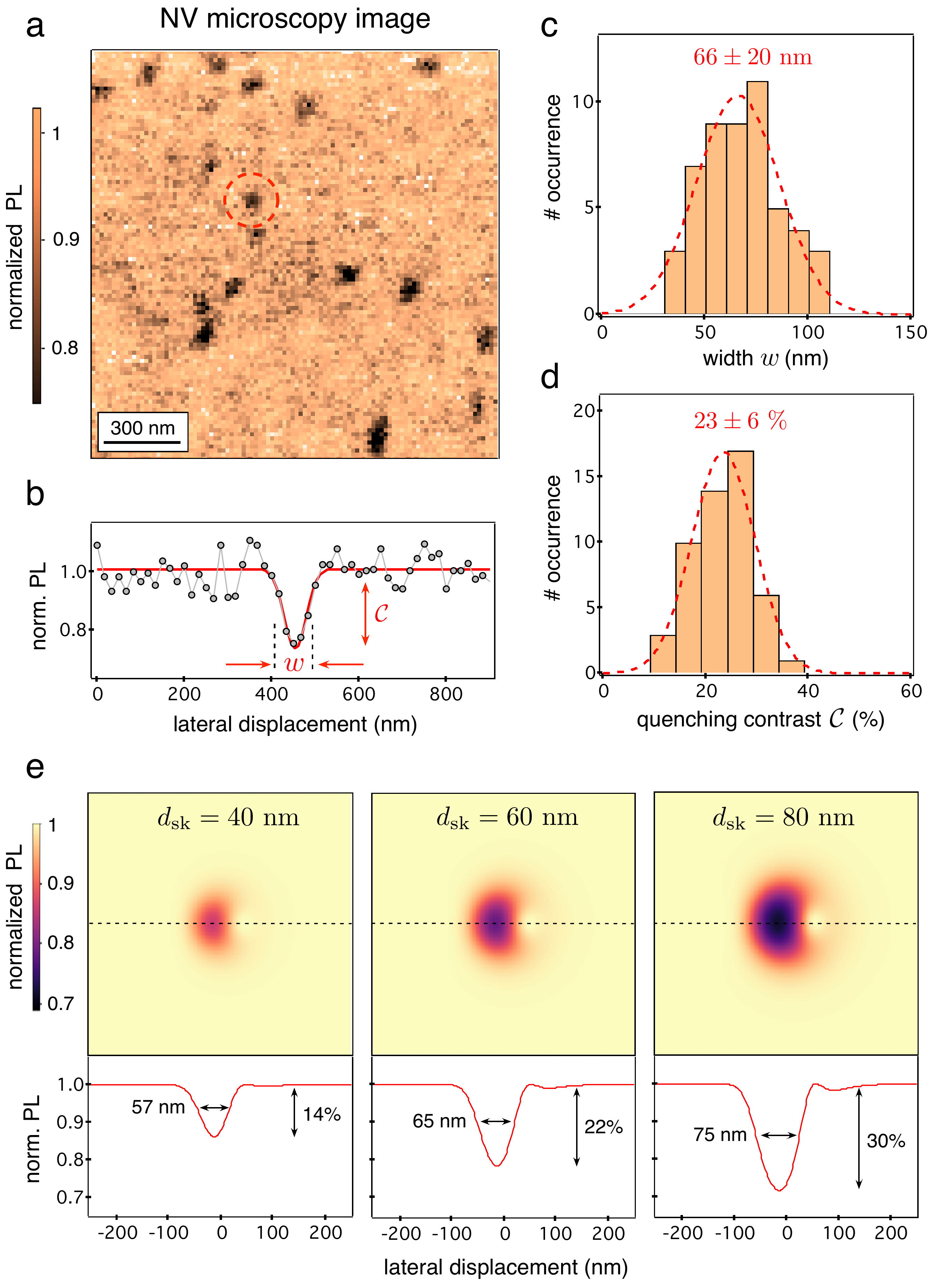}
\caption{{\bf (a)} NV magnetometry image recorded in photoluminescence (PL) quenching mode above the optimized exchange-biased multilayer stack at room temperature and zero external magnetic field. In these experiments, the distance between the NV sensor and the ferromagnetic layer is $\sim 65$~nm. {\bf (b)} Line profile across the skyrmion marked with the dashed circle in (a). The profile is taken along the fast scan direction (horizontal) and the red solid line is a Gaussian fit, leading to a full width at half maximum $w=58\pm8$~nm and a PL quenching contrast $\mathcal{C}=27\pm2\%$. {\bf (c,d)} Statistical distributions of (c) the width $w$ and (d) the quenching contrast $\mathcal{C}$, inferred from measurements over a set of 51 isolated skyrmions. The dashed solid lines are data fitting with normal distributions. The values written on top of the histograms correspond to the mean value $\pm$ standard deviation of the distributions. {\bf (e)} Simulations of the PL quenching image resulting from the stray field produced by skyrmions with diameters $d_{\rm sk}=40$~nm, $60$~nm and $80$~nm (from left to right). Line profiles taken across the skyrmions (black dashed lines) are displayed below the PL maps, indicating the width and the quenching contrast.
}
\label{Fig1}
\end{figure}

\begin{figure*}[t]
\includegraphics[width = 18cm]{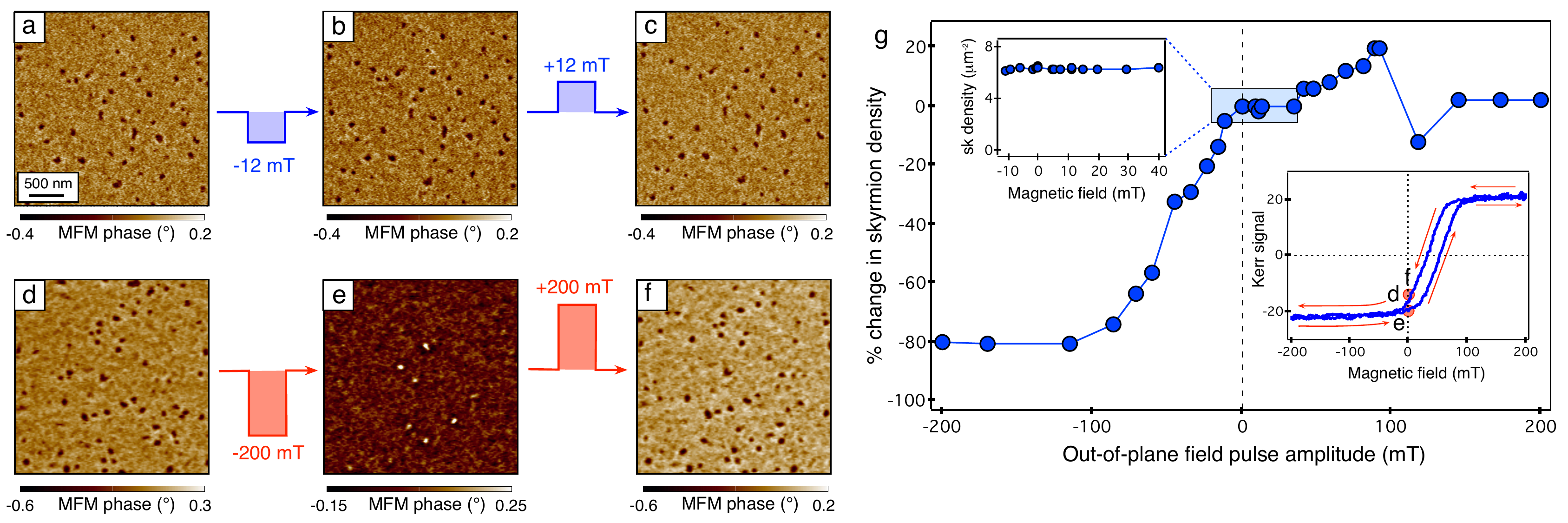}
\caption{{\bf (a-c)} MFM images recorded at zero external magnetic field before (a) and after applying a few-seconds out-of-plane field pulse of (b) $-12$~mT and (c) $+12$~mT. {\bf (d-f)} Same experiment before (d) and after applying a field pulse of (e) $-200$~mT and (f) $+200$~mT. {\bf (g)} Percent change in skyrmion density as a function of the magnetic field pulse amplitude. The top left inset is a zoom on weak magnetic field amplitudes. The bottom right inset is the hysteresis loop of the exchange-biased multilayer stack measured by recording the magneto-optical Kerr rotation as a function of out-of-plane magnetic field. The letters (d,e,f) refer to the MFM images.}
\end{figure*}

 Figure 2a shows a typical PL quenching image recorded at zero field above the optimized exchange-biased multilayer stack. Dark PL spots result from the stray field produced by isolated magnetic skyrmions. This observation confirms unambiguously the zero-field stability of skyrmions in the exchange-biased stack under ambient conditions. The full-width at half maximum $w$ and the contrast $\mathcal{C}$ of the dark PL spots were inferred by fitting line profiles with a Gaussian function for a set of 51 isolated skyrmions [Fig. 2b]. The resulting statistics are shown in Figure 2(c-d). Data fitting with a normal distribution leads to a mean value of the width $\bar{w}=66\pm 1$~nm, with a standard deviation $\sigma_w=20\pm2$~nm [Fig. 2c]. For the PL quenching contrast, which is linked to the stray field amplitude, a mean value $\bar{\mathcal{C}}=23\pm 1\%$ is obtained with a standard deviation $\sigma_{\mathcal{C}}=6\pm1$~nm [Fig. 2d]. In order to extract informations on the actual skyrmion size from these experimental results, simulations of the PL quenching image were performed for various skyrmion diameters. To this end, the skyrmion profile was simply modeled by a $360^{\circ}$ N\'eel domain wall with a left-handed chirality~\cite{PhysRevLett.114.177203} and a characteristic domain wall width $\sqrt{A/K_{\rm eff}}$, where $A$ is the exchange constant and $K_{\rm eff}=K-\mu_0 M_s^2 /2$ denotes the effective magnetic anisotropy. The exchange constant was fixed to $A=6$~pJ/m, a value in line with measurements in permalloy ultrathin films~\cite{Nembach2015}. The stray field distribution was first calculated at the flying distance of the NV sensor $z_{\rm NV}\approx 65$~nm for different skyrmion diameters~$d_{\rm sk}$. The resulting PL quenching images were then simulated by using the model of magnetic-field-dependent photodynamics of the NV defect described in Ref.~\cite{Tetienne2012}. Typical simulations of PL quenching images are shown in Figure 2e. Note that the slight asymmetric shape of the simulated images results from the orientation of the NV defect quantization axis, which is characterized by the polar coordinates ($\theta\sim 54^{\circ},\phi\sim 0^{\circ})$ in the $(x,y,z)$ reference frame. A fair agreement with experimental results is obtained for $d_{\rm sk}=60$~nm, both for the measured mean values of the width and PL quenching contrast. We can thus conclude that skyrmions with diameters around $60$~nm are stabilized at zero field in our optimized exchange-biased sample, corresponding to a reduction of skyrmion diameter by one order of magnitude compared to previous reports~\cite{NanoLett2018}. 

We note that smaller skyrmions, down to $\sim10$~nm diameter, have been recently observed at zero field in ferrimagnetic thin films close to the magnetization compensation at room temperature. However, such skyrmions were found as metastable states, with lifetimes limited to few hours~\cite{Caretta2018}. More generally, stable, sub-100~nm skyrmions have been obtained at room temperature by using multiple repetitions of magnetic multilayers, in which the thermal stability is enhanced by the large sample thickness~\cite{Fert2017,RevueSk}. However, skyrmion stabilization always requires to apply external magnetic fields in such multilayer stacks. Our results therefore demonstrate the potential of exchange-biased multilayer stacks to stabilize small-sized skyrmions at zero external magnetic field in ultrathin magnetic films. In such samples, exchange bias might enhance the skyrmion stability, a behavior already pointed out for the magnetic moment of magnetic nanoparticles~\cite{Nogues2003,EPL2009} or nanomagnets in MRAM cells~\cite{Prejbeanu2007}.  

 In order to characterize the impact of external magnetic field perturbations on the skyrmion stability, the magnetic landscape was systematically measured with MFM before and after applying a few-seconds pulse of out-of-plane magnetic field $B_z$. As an example, Figure~3(a-c) shows MFM images recorded at zero field before [Fig.~3a] and after applying a field pulse of either $-12$~mT [Fig.~3b] or $+12$~mT [Fig.~3c]. For such field pulse amplitudes, the skyrmion pattern is not modified. The impact of larger fields was analyzed by measuring variations of the skyrmion density. As shown in Figure 3g, the skyrmion density is little affected by field pulses with amplitudes lying between $-12$ and $+40$~mT.  For larger fields, a different behavior is observed depending on the pulse polarity. While the skyrmion density drops for negative field pulses below $-20$~mT, only small variations are observed for positive magnetic fields up to $+200$~mT. Negative magnetic fields tend to make skyrmion smaller and thus more unstable, possibly leading to their annihilation, which results in an overall reduction of the skyrmion density. Conversely, positive magnetic fields tend to enlarge magnetic skyrmions, thus increasing their stability, or leading to the nucleation of new skyrmions. An example of skyrmion patterns recorded at zero magnetic field before and after applying a large magnetic field pulse of $-200$~mT are shown in Figure 3~(d-e). A significant decrease of the skyrmion density is observed together with a reversed MFM contrast, in qualitative agreement with the lower Kerr signal evidenced in the hysteresis loop [see inset in Fig.~3g]. However, most skyrmions of the initial state can be recovered at zero field after saturation by applying  a large positive magnetic field pulse of $+200$~mT [Fig. 3f]. The magnetic skyrmions are therefore imprinted in the material, most likely through pinning effects induced by thickness variations in the multilayer stack~\cite{gross2018}. If an external magnetic field perturbs the domain structure, the information can be recovered after applying a large magnetic field pulse.\\
   
In this paper, we have shown that magnetic skyrmions with diameters around $60$~nm can be stabilized at room temperature and zero magnetic field in an optimized exchange-biased Pt/Co/NiFe/IrMn multilayer stack. In such samples, exchange bias provides an immunity of the skyrmion spin texture to moderate external magnetic field, in the tens of mT range, which is an important feature for applications as memory devices. These results establish exchange-biased multilayer stacks as a promising platform towards the effective realization of memory and logic devices based on the manipulation of topological spin textures.\\

\noindent {\it Acknowledgements:}  The authors thank A. Kovacs for his help with the acquisition of the STEM data and W. Akhtar for NV magnetometry experiments. This research has received funding from the DARPA TEE program, the European Research Council (ERC) under the European Union's Horizon 2020 research and innovation programme under grant agreements 639802 ({\sc Imagine}) and 856538 (3D MAGiC), the French Agence Nationale de la Recherche through the project {\sc Topsky} (ANR-17-CE24-0025) and {\sc Skylogic} (ANR-17-CE24-0045). A.~F. acknowledges financial support from the EU Horizon 2020 research and innovation programme under the Marie Sklodowska-Curie grant agreement 846597 ({\sc Dimaf}).

\bibliography{Sk}
\end{document}